\documentclass[conference,compsoc]{IEEEtran}
\ifCLASSOPTIONcompsoc
  \usepackage[nocompress]{cite}
\else
  \usepackage{cite}
\fi
\ifCLASSINFOpdf
\else
\fi
\usepackage{url}
\usepackage{tikz}
\usepackage{amsmath}
\usepackage{filecontents}
\usepackage{xspace}
\usepackage{hyperref} 
\usepackage{framed}
\usepackage{dirtytalk}
\usepackage{csquotes}
\usepackage{soul}

\definecolor{formalshade}{rgb}{0.96,0.96,0.96}
\definecolor{side}{rgb}{0.6,0.6,0.6}
\usepackage[strict]{changepage}
\usepackage{fancyhdr}
\usepackage[shortlabels]{enumitem}

\usepackage[table,xcdraw]{xcolor}         
\usepackage{wasysym} 
\usepackage{fancyhdr}
\usepackage{tabularx, booktabs} 
\usepackage{multirow, makecell}

\usepackage{
  color,
  xcolor,
  float,
  epsfig,
  graphics,
  graphicx,
}

\usepackage[color=cyan!30, textsize=small, disable]{todonotes}
\NewDocumentCommand{\WU}{o m}{%
  \IfNoValueTF{#1}%
    {\todo[color=pink!70]{WU: #2}}
    {\todo[inline, color=pink!70]{WU: #2}}
}

\definecolor{mygreen}{HTML}{A7D7A7} 
\NewDocumentCommand{\W}{o m}{%
  \IfNoValueTF{#1}%
    {\todo[color=mygreen!80]{WU: #2}}
    {\todo[inline, color=mygreen!80]{WU: #2}}
}

\NewDocumentCommand{\DZ}{o m}{%
  \IfNoValueTF{#1}%
    {\todo[color=cyan!30]{DZ: #2}}
    {\todo[inline, color=cyan!30]{DZ: #2}}
}

\NewDocumentCommand{\YZ}{o m}{%
  \IfNoValueTF{#1}%
    {\todo[color=yellow!30]{YZ: #2}}
    {\todo[inline, color=yellow!30]{YZ: #2}}
}

\newcommand{\longsay}[2]{%
 \def\FrameCommand{%
 \hspace{3pt}%
 {\color{side}\vrule width 2pt}%
 {\color{formalshade}\vrule width 4pt}%
 \colorbox{formalshade}%
 }%
 \MakeFramed{\advance\hsize-\width\FrameRestore}%
 \noindent\hspace{-4.55pt}
 \begin{adjustwidth}{}{7pt}%
 \vspace{1pt}%
  \say{\textit{#1}} %
 \vspace{0.5pt}%
 \end{adjustwidth}%
 \endMakeFramed%
}

\usepackage{tcolorbox}

\newcounter{adviceCounter}

\newcounter{futuredirectionsCounter}

\newcounter{needsCounter}

\newcommand{\n}{23\xspace}

\newcommand{\HCTM}{human-centered threat modeling\xspace}

\renewcommand{\paragraph}[1]{\textbf{#1}}

\usepackage[normalem]{ulem}

\usepackage{soul}
\begin{document}
%
\title{Human-Centered Threat Modeling in Practice: Lessons, Challenges, and Paths Forward}

\author{

\IEEEauthorblockN{Warda Usman}
\textit{Brigham Young University            }
\and
\IEEEauthorblockN{Yixin Zou}
\textit{Max Planck Institute for Security and Privacy}
\and
\IEEEauthorblockN{Daniel Zappala}
\textit{Brigham Young University}
}

%


\maketitle

\begin{abstract}
 
Human-centered threat modeling (HCTM) is an emerging area within security and privacy research that focuses on how people define and navigate threats in various social, cultural, and technological contexts. While researchers increasingly approach threat modeling from a human-centered perspective, little is known about how they prepare for and engage with HCTM
in practice.
In this work, we conduct 23 semi-structured interviews with researchers to 
examine the state of HCTM, including how researchers design studies, elicit threats, and navigate values, constraints, and long-term goals.
We find that HCTM is not a prescriptive process but a set of evolving practices shaped by relationships with participants, disciplinary backgrounds, and institutional structures. 
Researchers approach threat modeling through sustained groundwork and participant-centered inquiry, guided by values such as care, justice, and autonomy. They also face challenges including emotional strain, ethical dilemmas, and structural barriers that complicate efforts to translate findings into real-world impact.
We conclude by identifying opportunities to advance HCTM through shared infrastructure, broader recognition of diverse contributions, and stronger mechanisms for translating findings into policy, design, and societal change.

\end{abstract}


%
\IEEEpeerreviewmaketitle

\section{Introduction}

Security and privacy research has long recognized that building effective protections requires understanding users’ behaviors, needs, and   practices \cite{adams1999users, cranor2008framework}. While much of the field  historically focused on securing technical systems, researchers increasingly recognize that addressing security and privacy risks requires engaging not only with technical systems, but with how people perceive, experience, and manage threats in their everyday lives. 

Researchers have begun using \HCTM (HCTM) to better identify how security and privacy can meet human needs, focusing on \textit{threats to people} rather than systems and aiming to surface harms as they are lived within specific social, cultural, and technological contexts \cite{usman2024sok}. By centering people's lived experiences, HCTM enables researchers to surface overlooked threats and guides the development of technical, educational, policy, or societal solutions that are more realistic, effective, and equitable. 
There has been a growing body of research modeling threats across a range of contexts. Studies have focused on threats to specific communities \cite{frik2019privacy,mcnaney2022exploring, slupska2022they, lerner2020privacy, daffalla2021defensive}, threats from particular technologies \cite{lau2018alexa, munyendo2022desperate, gallardo2023speculative}, individual experiences of harm \cite{wei2023there, take2022feels, stephenson2023s}, and risks to populations in particular locations \cite{sambasivan2019they, ahmed2017digital, wilkinson2022many}. Together, this work reflects a wide-ranging effort to understand how the field of usable security and privacy can better meet a diverse array of human needs.

Recognizing the need to bring structure to this emerging area, recent work has begun to formalize the study of human-centered threat modeling. Usman and Zappala \cite{usman2024sok} systematized knowledge from 78 papers that conduct HCTM. Their SoK introduces a formal definition of HCTM and proposes a framework structured around four components---context, threats, protective strategies, and reflection. Their work also offers practical guidance for conducting HCTM studies, identifies how HCTM differs from traditional systems-based threat modeling, and illustrates how deep engagement with a community can help achieve researcher goals to center human safety.

While this SoK significantly advanced our understanding of HCTM, 
it could only surface what was reported in publications. As a result, we still know little about the behind-the-scenes practices that researchers engage in and the challenges that force researchers to forgo pursuing certain efforts. For example, the HCTM framework \cite{usman2024sok} includes groundwork as a preparatory step and emphasizes that context influences the HCTM process. However, we don't know the specific challenges researchers face when conducting such groundwork or why researchers choose to focus on one context over another.
Because the process of building human-centered threat models depends so heavily on researchers' situated judgments and adaptations \cite{usman2024sok}, understanding researcher experiences and viewpoints can offer important lessons about HCTM for novice researchers, identify challenges that hinder greater impact, and provide insights for the field to become more effective.

To better understand the behind-the-scenes practice of HCTM, we conducted interviews with 23 researchers. Our interviews covered a wide range of topics, spanning current practices (groundwork, how threats are elicited), challenges, vision (long-term goals, making an impact), motivation, and advice to new researchers. These topics evolved during the early stages of interviews, based on themes that emerged, and were all carefully centered on HCTM rather than broader usable security and privacy research.
This afforded us a rare opportunity to reflect on the field’s current trajectory and identify opportunities for strengthening the impact of HCTM as researchers continue to seek solutions that meet human needs.

Contributions of this paper include:

\begin{itemize}

    \item We analyze researcher motivations and long-term goals for pursuing HCTM, highlighting how deeply held values such as care, justice, autonomy, and safety shape research choices.
    
    \item We surface current practices for designing and conducting HCTM studies, centered on the importance of sustained groundwork and best practices for eliciting threats through broad participant-centered inquiry.

    \item We identify emotional and practical challenges researchers face conducting HCTM studies, as well as structural barriers that limit the translation of findings into action. 

    \item We investigate how researchers envision HCTM making greater impacts through stronger engagements with industry, policy makers, and communities. Researchers also identify needed changes to their research culture, including expanding what counts as a valuable contribution beyond publications and fostering new forms of collective infrastructure to sustain work over time.

\end{itemize}

We conclude by discussing lessons learned and future directions for HCTM, including building stronger bridges to industry and policy and embracing pluralism in methods and evaluation standards. We also highlight the political constraints that U.S. researchers will need to navigate in order to continue obtaining funding for HCTM research.

\section{Background and Related Work}

 \subsection{Human-Centered Threat Modeling}
HCTM is the process of identifying how individuals or communities perceive, navigate, and manage risks to themselves. It centers on people's lived experiences and is grounded in understanding the social, cultural, and technological contexts in which they operate.
Usman and Zappala \cite{usman2024sok} organize HCTM into four interconnected components: \textit{context}, the environments in which risks emerge; \textit{threats}, the concerns or harms participants identify; \textit{protective strategies}, the actions taken to mitigate those risks; and \textit{reflection}, the process of evaluating with participants what strategies work, what barriers exist, and what changes are needed.

While HCTM research spans a wide range of topics, we highlight several areas of focus to illustrate its breadth. As an example of research focused on a particular threat, researchers have extensively studied online hate and harassment. An SoK from Thomas et al. \cite{thomas2021sok} offers a taxonomy of attack types, prevalence estimates, and risk factors, and proposes community-oriented responses.  As an example of research focusing on populations, researchers have studied at-risk groups, such as activists \cite{lerner2020privacy, daffalla2021defensive}, journalists \cite{goyal2022you}, members of the LGBTQ+ community \cite{lerner2020privacy, scheuerman2018safe}, or refugees and immigrants \cite{simko2018computer, tabassum2025privacy, usman2025security}. An SoK by from Warford et al. \cite{warford2022sok} unifies this research by developing a set of risk factors, illustrating how these factors put different categories of groups at increased risk. They also identify a wide range of protective practices and barriers to protective practices. 
As an example of research focusing on a particular technology, researchers have studied privacy perceptions of IoT devices in a variety of contexts, such as smart speakers \cite{morethansmart, huang2020amazon, lau2018alexa, meng2021owning}, smart homes ~\cite{sun2021child, bystanders, tabassum2019don, zheng2018user}, and smart cities \cite{emami2023understanding}.

HCTM builds on the tradition of systems threat modeling, a foundational practice in security engineering aimed at identifying and mitigating vulnerabilities in technical systems \cite{shostack2014threat}. 
In systems threat modeling, threats are often generalizable and cataloged in frameworks 

such as STRIDE \cite{MicrosoftCorporation2005} or attack trees \cite{attacktrees}, 
with security engineers viewed as the domain experts responsible for defense.
In contrast, HCTM reorients the process around people rather than systems \cite{usman2024sok}.  People come with contexts that are shaped by their social identities and lived experiences. Likewise, threats are situated, not standardized, and encompass technical, physical, emotional, relational, or societal harms \cite{scheuerman2021framework}. Here, participants are the experts on their own risks, and researchers serve as partners in surfacing those threats. 
These differences mean researchers must engage deeply with context, adapt methods, and navigate ethical and methodological challenges. 

Understanding how researchers carry out this work offers crucial insight into how HCTM is practiced and how it can continue to evolve.

\subsection{Meta-research in Usable Security and Privacy}

While human-centered security research often focuses on specific technologies or populations, a smaller body of work has turned the lens inward to study how research itself is  conducted, reported, and understood within the community. The aim of meta-research is to strengthen practices related to methods, reporting, reproducibility, evaluation, and incentives \cite{ioannidis2015meta, le2023analyzing, hasegawa2024weird}.
This kind of reflection contributes to broader efforts to strengthen the \textit{\say{science of security,}} a movement that calls for a more systematic and rigorous foundation for security research \cite{herley2017sok}. 

Herley and van Oorschot caution against importing scientific norms from other fields without examining the unique goals and constraints of security research, noting a lack of consensus around evidence, progress, and epistemology  in the field of security \cite{herley2017sok}. In contrast, Spring et al. \cite{spring2017practicing} contend that security already operates as a science, not one grounded in universal laws or falsifiability, but in structured observation, mechanistic explanation, and pluralistic inquiry.

In usable security and privacy, this reflexive turn has taken several forms. Hasegawa et al. reviewed participant samples across published studies, revealing a strong WEIRD bias that limits generalizability and reproducibility \cite{hasegawa2024weird}. Jacobs and McDaniel conducted a qualitative analysis of user-centered research, identifying key themes, methodological trends, and gaps in how end-users are represented \cite{jacobs2022survey}.
Klemmer et al. examined transparency in study design and reporting, revealing a gap between researchers’ values and what venues reward \cite{klemmer2025transparency}. Bellini et al. combined literature review and oral histories to surface the often-invisible labor of mitigating digital-safety risks for both participants and researchers \cite{bellini2024sok}. Distler et al. reviewed how risk is modeled in user studies, finding wide variation in methods \cite{distler2021systematic}. 
Suray et al. analyzed \say{future work} statements in SOUPS papers, tracing whether authors followed through, and offering insight into how research agendas evolve over time \cite{suray2024future}.

\subsection{Social, Moral, and Political Values in Science}
\label{sec:value}

Debate over the role of values in science has lasted for decades \cite{douglas2009science}. The idea that science is and should be value-free rests on the argument that science objectively pursues truth using neutrality and impartiality, with moral autonomy from society \cite{lacey2005science}.  Douglas, on the other hand, argues that scientists must consider the consequences of their work and therefore can't maintain a value-free ideal \cite{douglas2009science}. She believes social, moral, and political values are an essential part of scientific reasoning. Her work leads not to a rejection of scientific objectivity, but instead a framework for how to use values in science.\footnote{Many other critiques of value-free science have been made, notably from a feminist perspective \cite{harding2019objectivity}.} Brown summarizes the decades of debate by stating that there is \say{\textit{near consensus among philosophers of science ... that the ideal of value-free science is untenable, and that science not only is, but normatively must be, value-laden in some respect.}} \cite{brown2024values}

Within computer science, Nissenbaum has argued that computer systems embody values \cite{nissenbaum2001computer} and that those who design computer systems have a responsibility to take values into consideration, enumerating values such as ``liberty, justice, enlightenment, privacy, security, friendship, comfort, trust, autonomy, and sustenance.'' \cite{flanagan2008embodying}. Likewise, CS educators have argued for incorporating social values into ethics courses \cite{ferreira2021deep} and into the undergraduate curriculum as a whole \cite{moudgalya2024need,lin2022cs}.

When science intersects with political values and public policy, it has led to debates about the role of science in politics. This is particularly noted within the areas of climate science \cite{van2024challenging}, public health \cite{coggon2022public, claessens2021science}, and various areas of CS research, including disinformation, AI, and tech harms \cite{moran2025privacy}. The current political climate in the U.S., with numerous cancellations of science and education grants, particularly aimed at ending diversity, equity, and inclusion  efforts, can be framed as part of a larger ``attack'' on science from some political interests \cite{attacksonscience}. This has led to calls for collective action from the university community \cite{pfeifer2022combatting}, with some arguing that science has always been political \cite{thorp2020science}, and scientists themselves becoming  political activists \cite{fuentes2024scientists}.

Our work demonstrates that researchers engaged in HCTM are motivated by social and moral values, adding to the complex picture of scientists pursuing research that may conflict with efforts to end government funding for work that aims to help vulnerable or at-risk populations \cite{singh2025epa, nyt2025fund}.

\section{Methods}
\label{sec:methods}

In this study, we explored how researchers conduct HCTM in practice, their challenges and motivations,  and the future they envision for the field. To this end, we conducted \n interviews with researchers actively engaged in HCTM.
\subsection{Selection Criteria and Recruitment}

To identify researchers actively engaged in HCTM, we began by using an open dataset\footnote{\href{https://github.com/Usable-Security-and-Privacy-Lab/IEEESP25-SoK/blob/main/corpus.md}{Dataset by Usman and Zappala.}}
of HCTM papers from Usman and Zappala
\cite{usman2024sok}. 
We supplemented this corpus by searching for additional recent papers published on HCTM at major S\&P and HCI venues {(IEEE S\&P, USENIX Security, SOUPS, CCS, NDSS, PETS, CHI, CSCW)}. Our operational definition of HCTM incuded  research examining how individuals perceive and respond to threats affecting their safety, privacy, or security \cite{usman2024sok}.  

While this approach helped identify an initial pool, we found that many authors from the original dataset were no longer actively working in HCTM. To focus on researchers currently engaged in this area, we shifted to identifying individuals who had recently published related work at these venues.

Researchers were eligible if they had authored at least one paper falling under our operational definition of HCTM, even if their work did not explicitly refer to it as such. To recruit participants for our study, we contacted them via email, inviting them to voluntarily contribute their perspectives and insights as participants.

\subsection{Participants}

To protect participant privacy in the small, well-connected  research community, we report only aggregated demographics. Our sample comprised \n researchers actively engaged in human-centered threat modeling, most having published multiple HCTM papers in the venues listed above.
Participants were based in the United States, United Kingdom, Germany, and Canada, and represented a diverse range of institutions, including R1 universities, liberal arts colleges, government agencies, and industry research labs.
Participants had varying levels of experience: eight were senior researchers (including full professors and industry leads), nine were mid-career (primarily assistant and associate professors), and six were junior researchers (PhD students or early-career postdocs).
Roles included one instructor, five assistant professors, three associate professors, two full professors, five PhD students or candidates, two postdoctoral researchers, and five participants in research scientist, practitioner, or engineering roles. 
Most participants (18) were based at academic institutions, with the remaining affiliated with government labs, research institutes, or industry.
The interviewees collectively published 1106 (median = 22) papers over their careers, including 246 on human-centered threat modeling (median = 8).

\subsection{Interviews}

We conducted semi-structured interviews to gather insights into their experiences, practices, and challenges. 
While we began with a guiding set of questions, our interview protocol evolved as themes emerged and our understanding of the space deepened. Early interviews focused on how researchers conduct HCTM (such as their study designs, threat elicitation methods, and community engagement), because we were interested in surfacing methodological patterns across projects. However, as researchers brought up their values, motivations, and goals that guided their work, we added questions on these topics after the third interview.  We likewise added questions about next steps to elicit opinions about where the field is going, partly informed by our engagement with the field (\ref{sec:positionality}), partly by keen participant interest in the subject.

To set the context for the interview, we began by explaining the study’s purpose, provided our definition of HCTM, and briefly discussed their work on HCTM. We asked participants if they agreed with our categorization of their work to verify their eligibility and begin the conversation.

During the interview, we explored their research processes, approaches to groundwork, challenges, long-term goals, values, and perspectives on the future of the field of HCTM.  
To scope the discussion, we began by focusing on a specific paper or set of papers from their body of work, which we had selected in advance. This allowed us to ask detailed questions about that particular project, such as their processes, approaches to groundwork, and the challenges they encountered.
For broader topics, such as their values, long-term goals, and perspectives on the future of the  field, we expanded the discussion beyond individual projects. These questions were designed to capture their overall outlook on their work and the direction of the field.

We conducted interviews via Zoom, with audio recordings captured through the platform and transcriptions generated using \texttt{Otter.ai}. The first author reviewed all transcripts for accuracy and consistency with the recordings. We informed participants in the consent form that we would use \texttt{Otter.ai} for transcription.  We did not offer compensation to participants. Interviews lasted an average of 39 minutes.

\subsection{Data Analysis}
To analyze our data, we employed thematic analysis \cite{braun2006using}, using a hybrid approach combining deductive and inductive coding. We first familiarized ourselves with the transcripts, and created seven primary categories based on our research objectives: groundwork, how to elicit threats, challenges, long-term goals, values, advice, and next steps. These categories formed the initial structure of our codebook. Within each of these pre-defined categories, we then applied inductive coding to the entire transcript. This involved reading through the whole transcript as a cohesive data unit and identifying emergent themes, patterns, and concepts. This hybrid approach allowed us to maintain focus on our research objectives while remaining open to unexpected insights and nuanced themes that emerged from the data itself \cite{tavory2014abductive, coffey1996making}.
Two researchers jointly coded all the transcripts and any disagreements were resolved via on-the-spot discussions. 
We identified themes iteratively throughout the interview and coding process, writing memos and discussing potential themes together. We regularly revised and synthesized themes to capture insights we felt would be valuable to researchers engaged in HCTM.

\subsection{Member Checking}

\label{sec:mem-checking}
After completing the draft, we shared the paper and a short summary\footnote{All our supplemental materials, including the summary document are available \href{https://osf.io/n623d/overview?view_only=6706c79e929441f3a8c66fd0ef20fc05}{here}.}
with all participants to validate our interpretations and representation of their interviews. 
Following best practices for member checking \cite{mckim2023meaningful}, we asked whether the findings represented their experiences accurately and whether there was anything they would like to add, remove, or revise. We provide the invitation email and questions in Appendix \ref{member_checking}. Feedback from this process informed several clarifications and minor refinements to the final paper.

\subsection{Researcher Positionality and Values}
\label{sec:positionality}
We approach this study as researchers situated within the human-centered security and privacy community. Our interest in HCTM stems from our own experiences conducting qualitative security and privacy research with marginalized populations, and from a broader commitment to understanding how security research can be made more inclusive, context-sensitive, and socially responsive. We also have conducted systems research and threat modeling for systems.
We view HCTM as a sociotechnical process shaped by the context of participants as well as the relationships between researchers and the communities they study.

We acknowledge that our dual roles as interviewers and as fellow researchers in this space may have influenced how participants shared their experiences. We took care to frame the interviews as collaborative and exploratory.
We see this work as part of an ongoing conversation about how the field defines its goals and models threats in ways that reflect people's lived experiences.

\subsection{Limitations}
The perspectives included here reflect the experiences of researchers who regularly publish HCTM-focused papers in leading security and privacy venues and who were willing to participate, but may not capture the full diversity of HCTM practices.
We aimed to include a diverse sample in terms of expertise, methodological approach, and experience level. However, our participant pool was limited to individuals based in North America and Europe, despite efforts to recruit researchers from other regions. Our participant pool was also driven in part by the personal connections of the authors to other researchers in the space. 

Additionally, our study is based on self-reported accounts, which reflect how participants remembered, framed, and interpreted their own work. These narratives may not fully align with how their practices appear in publications or how others in the field perceive them. Finally, our positionality as researchers within the HCTM space shaped the questions we asked and how we interpreted participant responses. We acknowledge that other researchers may draw different insights from the same conversations.

Despite these limitations, we believe this work offers a valuable window into how HCTM is currently practiced and provides a foundation for future research to examine, critique, and expand this emerging area.

\section{Findings}

Our findings span four topics, covering \textit{motivation and values} (why the work is pursued), \textit{work behind the scenes} (how the work is conducted), \textit{challenges} (what gets in the way of the work), and \textit{next steps} (how impact can be increased).

\subsection{Motivation and Values}

We talked with researchers about how they chose a threat-modeling research project, as well as their long-term goals. These topics helped us to uncover their motivations for engaging in HCTM, which were grounded in their personal values.
While not all participants directly articulated how their values influenced their research process, for many, values shaped not only the motivation for doing HCTM but also the trajectory of the work.

Researchers who regularly engaged in community-centered threat modeling were strongly motivated by values such as caring, humility, fairness, and social justice. Researchers who took a more technology-centered approach typically focused on agency, autonomy, privacy, or safety. We emphasize that not all values were universally held; rather we report the totality of values researchers expressed.

\subsubsection{Caring about people}

Those engaged in community-based research framed their work around caring, hoping they could benefit a specific community:

\longsay{It comes from this  underlying assumption that we care about the people that we're researching and we want the research to benefit them as much, if not more so, than it benefits us, right? And not coming from it with a perspective of like, well, I'm just doing this because it's trendy, because it's more popular now.}

One researcher saw beauty in understanding how people interact:

\longsay{I want them to eventually come around to seeing like, the ways that their norms and behaviors are, like, really beautiful, like, just gorgeous, interesting ways of connecting, setting boundaries,  building community, and that they are  valued and have led to breakthroughs in our conception of  what privacy even is.}

Finding a community they could help sometimes started with students and their interests. One senior researcher explained that their lab tends to attract students with a strong commitment to a community, which organically shapes the lab’s research agenda:
\longsay{We’ve established this lab that uses certain methods and asks certain types of questions. And we tend to attract students who really care about supporting different types of communities. }

In other cases, the decision to pursue a topic stemmed from long-standing personal interest or emotional investment in an issue. For example, having a family member with a disability led to one researcher's interest and subsequent work with that community. 
Further, researchers reported that they tend to avoid engaging with communities they are not part of or with which they lack familiarity.

\subsubsection{Humility and respect}

Researchers expressed values centered around protecting communities they worked with. They
mentioned humility, recognizing that they may not be a part of the community they study and may not fully understand
their concerns or the harms community members experience. They stressed centering participant voices and being careful to separate their own views from what participants say. Researchers discussed striving for respect, to ensure that they accurately represented the perspectives and behaviors of participants:

\longsay{I feel like that the populations that I study have had bad experience with researchers because they're used to researchers coming to them, studying them as outsiders, from the perspective of outsiders, and not bothering to understand their values and how those values influence their behaviors...it's so off putting, like, I don't want to be studied so that someone can call me weird, right?} 

Similarly, another participant said that \say{\textit{one person also told me that they don't like to be studied like exotic animals in a zoo}}.  Likewise, participants expressed that participants deserve dignity and have a right to anonymity. One participant expressed a \say{\textit{principle of participation,}} meaning that they would strive to hire members of the community on their research team. Another expressed a desire to help participants understand the valuable contributions they had made to research. Finally, participants mentioned following standard ethical practices in the research community, including ensuring informed consent, maintaining confidentiality, transparently communicating
how participant data will be used, and avoiding research that could lead to harm.

\subsubsection{Social justice and equity}

Some participants expressed a desire to work toward social justice or equity, in line with prior work discussing embedding justice elements in research \cite{BhaleraoEthics}. One participant explained:

\longsay{I would see the world as sort of fundamentally a place with a lot of inequities. And if we want the world to be a safer place, I think the world also needs to be a more fair and just place. I think a lot of security, privacy, safety problems are a result of systems that value some people over others.}

Others described equity as ensuring that everyone's voice is represented equally, or that systems were designed so that they
work for everyone. Still others expressed a desire to avoid harm, especially for those with the least power in society.
A different set of participants spoke about wanting to  \say{\textit{make the world a better place}}.
Solving problems in society was seen as more important than publishing a paper. Helping people didn't necessarily equate to providing
a technical solution; in fact, often researchers acknowledged that technology might not necessarily help or might be part of the problem.
For some, they viewed their research as simply
helping to make sense of the world. Their commitment to justice involved working with groups that are often criminalized.  {One participant shared that their efforts to surface and mitigate harms experienced by such communities were sometimes met with skepticism, as if acknowledging those threats implied endorsing the group's actions. This pushback reveals how threat modeling is not only guided by researcher values, but also constrained by broader societal narratives about whose safety is seen as legitimate.} 
In contrast, some participants struggled with the notion that they might want to make the world a better place, indicating they didn't know whose
``version'' of a better place was more correct, or stating that they didn't want to push their vision on the rest of the world.

\subsubsection{Privacy and autonomy}

Naturally, researchers conducting HCTM shared values related to security, safety, and privacy. What was notable was how they conceptualized and expressed their privacy values. Some researchers had particularly strong conceptions of privacy, expressing the belief that privacy is a fundamental right, or that technology should implement privacy as a default. For example, one researcher noticed that Venmo’s default setting made all transactions public and described this design as “weird” from a privacy perspective. This observation motivated a study on social norms and default visibility in financial apps. 
Other researchers viewed themselves as making practical trade-offs between privacy, security, and usability, or not being dogmatic about it.

Participants linked privacy to agency and autonomy, expressing a desire to enable people to make decisions regarding their information. 
This was grounded in a value of respecting people's choices and withholding judgment if they thought or acted differently than the researcher.

Some researchers focused primarily on autonomy as a broader value, discussing a desire to make knowledge accessible outside of academia, so that people could be empowered.
Likewise, others connected autonomy and respect to ``democratizing tech'', consisting of openness, transparency, and accessibility.

\subsubsection{Tensions}

Participants relayed a number of tensions concerning their values. They acknowledged the impact their values had on their research, and some indicated that they could be listed in a positionality statement. However, others struggled with whether it was appropriate to mention them in a scientific paper, worrying that the community wouldn't accept papers that mentioned values. This reflects the tension between a values-free ideal of science versus a value-infused reality (\S\ref{sec:value}). Other researchers discussed that research based in social justice values, or research taking a critical or interpretivist perspective, was in the minority view in the security and privacy community. 
This could lead to negative peer reviews or questions from funding sources about the need
for this kind of work.

\subsection{Work Behind the Scenes}

We asked researchers how they approach human-centered threat modeling in practice. We highlight two key aspects of this work: the often-unreported groundwork that shapes how studies begin and evolve, and the strategies and decisions researchers make when eliciting threat models from participants. Together, these findings provide a view into how human-centered threat modeling is enacted day-to-day, beyond what appears in published papers.

\subsubsection{Groundwork}

\label{subsec:groundwork}

We talked with researchers about the types of groundwork they engaged in to inform their HCTM studies. This helped us explore an aspect of research practice often missing from published papers: the early-stage efforts researchers undertake to understand the communities, technologies, and threats they study. Participants described groundwork as essential to HCTM, not as a preliminary step, but as a continuous process that unfolds throughout the research and actively shapes how threat models are surfaced and understood. 

Researchers often framed their contributions through one of three lenses: a focus on a particular \textit{community} (e.g., immigrants, incarcerated people, women, older adults), a   \textit{technology} (e.g., AirTags and smart home devices), or a   \textit{threat} (e.g., harassment, image-based sexual abuse, and surveillance). {These orientations frequently intersected but broadly fell into two types of groundwork}: community-centered (starting from the lived experiences and needs of a population) and technology-centered (starting from the features, affordances, and constraints of specific systems). When researchers oriented their work around a threat, they typically grounded the study in a specific population, a particular technology, or both. 
Accordingly, we focus here on the two primary types of groundwork that emerged from our participants: \textit{community-centered groundwork} and \textit{technology-centered groundwork}.

Although presented separately here, community- and technology-centered groundwork often co-evolved. Understanding lived experiences frequently revealed technical questions to investigate, and hands-on technical analysis uncovered social dynamics that required community insight. Together, these intertwined practices illustrate how much of HCTM’s labor happens before formal data collection and continues on afterward.

\paragraph{Community-centered Groundwork.}
Researchers stressed that meaningful community involvement is critical for identifying threats rooted in lived experience, since without it, threat modeling risks becoming abstract or misaligned.
 
\longsay{I will want to connect with people in the community to check my understanding and also make sure that whatever I’m going to solve is actually a problem for the community. I never want to   parachute in and be like, \say{Aha, I think you have a problem},  and they’re like, \say{No, this is not a problem. Why are you here? Go away.}}{P10}

Participants described groundwork as unfolding along three interrelated dimensions. 
\textit{(1) Connecting with the community} involves initiating relationships and building trust. Researchers described doing so through local organizations, informal networks, or pre-research activities like workshops to establish access and credibility.
\textit{(2) Understanding the community} entails learning how communities define and experience threats, harms, and protections. Researchers engaged with public discourse, adopted local language, and drew on interdisciplinary knowledge to avoid imposing external assumptions.
\textit{(3) Working with the community} refers to ongoing collaboration during the research process. Participants described co-developing materials, sharing findings, and working with community insiders to ensure the research remained respectful, context-sensitive, and grounded in the community’s own ways of articulating risk.

While they did not explicitly use the term community-based participatory research (CBPR) \cite{satcher2005methods, hacker2013community}, 
we observed strong parallels in their emphasis on sustained, equitable engagement. 
Table \ref{tab:groundwork} in Appendix~\ref{sec:strategies} outlines the strategies participants shared for each dimension, along with example stories and how these practices align with CBRP principles. Some of these practices also intersect with Bellini et al.~\cite{bellini2024sok} as they recommend best practices for participant engagement when conducting digital safety research with 
at-risk groups.  
Our participants framed similar practices as critical for grounding threat modeling in lived experience and ensuring contextual accuracy.

\paragraph{Technology-centered groundwork.}
 Those studying specific technologies described a need to deeply understand the tools and practices. For HCTM, knowing what a technology is supposed to do is rarely enough. Researchers need to understand how it actually behaves, including what data it collects, how users interact with it, what assumptions are embedded in its design, and how it might exacerbate harm.

To do this, researchers engaged in various forms of groundwork. Some reviewed prior literature to identify what others had found about a particular system. Some bought or used the technology themselves, replicating typical user behaviors to explore the system’s surface-level functions as well as what happens in the background. This often meant going beyond the UI to examine data flows, privacy settings, logging behavior, or default permissions.

\longsay{It usually starts with a bunch of Google searches, but that only goes so far. So we end up buying a bunch of smart home devices to see how they actually behave, especially when it comes to stuff like   access controls. A lot of the info online is either outdated or very superficial, so testing things ourselves is the only way to really understand what’s going on.}

One researcher shared how their groundwork involved analyzing traffic data and domain rankings to understand the popularity of different modified versions of a widely used messaging app. They then conducted exploratory analysis on the most prominent variants, examining permissions, third-party libraries, and potential security risks.

Others described informally learning from the people around them, such as talking to friends, family, or people they met in everyday life to hear how technologies were being used and what kinds of problems or workarounds people had developed.  

\subsubsection{Eliciting threat models}
 
We asked researchers how they go about understanding the threat models of their participants, with the goal of surfacing both current practices and emerging best practices in HCTM.  While Usman and Zappala have identified the methods researchers use in their SoK by analyzing papers \cite{usman2024sok}, our study provides deeper insights into researchers’ epistemological reflections on how those methods are applied in practice. These reflections underscore that the quality of a threat modeling exercise depends not just on the method used, but on how the study is carried out including  how questions are framed, when they are introduced, and the context in which they are asked \cite{redmiles2017summary}.

Researchers emphasized that centering people is essential. Without engaging participants directly in the   process, researchers may miss critical aspects of their threat landscape and design solutions that fail to address real concerns.
\longsay{By selecting specific research projects, we sometimes do the threat-identifying part for users, but in doing so, we often forget that participants’ own views of which threats are important  may be different.}{}
Researchers also emphasized that the success of a HCTM exercise heavily depends on thorough groundwork (\S \ref{subsec:groundwork}). Further, rather than arriving with a fixed view of participants’ experiences, researchers treated threat modeling as an iterative process. They continuously refined their understanding through ongoing interaction, allowing participants’ concerns and definitions of harm to guide the direction of the work.  
Researchers often began with broad questions, allowing participants to define threats in their own terms. Direct questions about privacy or security tended to yield limited responses, as participants rarely used those terms, even when describing relevant behaviors. This aligns with prior work showing that people understand  privacy contextually and practically \cite{solove2021myth, nissenbaum2004privacy}.   One researcher reflected:
\longsay{So in a recent study, we did ask the question, \say{What does privacy mean to you?} And our participants said almost nothing.  But we had just spent 30 minutes discussing behaviors clearly tied to privacy. We got much more rich data when we talked to them about specific issues that their community was facing, and their behaviors in response to those issues.}{}    

To   elicit threat models, researchers often relied on a consistent set of keywords or \say{entry points}. Researchers frequently mentioned asking about \textit{cybercrime}, \textit{negative experiences}, \textit{uncomfortable situations}, \textit{challenges}, \textit{concerns}, and \textit{things participants didn’t like about a system}. They also asked about protective strategies, which offered insight into what participants felt the need to defend against. Some researchers used structured prompts based on prior work, such as taxonomies by Thomas et al. \cite{thomas2021sok} or Warford et al. \cite{warford2022sok}, to ensure important risks were not overlooked. 

One researcher emphasized that effective threat modeling goes beyond listing potential risks. What differs across populations is not only which threats exist, but how they are experienced, how frequently they occur, and the extent of harm they cause:

\longsay{Ideally, there’s already research that establishes the baseline -- what the concerns are, what the risks are. So for most of the groups we work with, the question isn’t really \textit{what} the risks are. We sort of already know that. It’s more about \textit{which} of those risks are more severe for them. Which ones are they more likely to face? Which ones are more prevalent?}{}

\subsection{Challenges}
We asked researchers about the challenges they face to better understand the   barriers that shape how human-centered threat modeling is done. Researchers had few answers to these challenges, indicating these may be areas where the community needs to organize solutions.
\subsubsection{Navigating researcher harms}
Researchers frequently  experience significant emotional and psychological strain in their work.
The research process requires a degree of vulnerability from researchers, which can be emotionally taxing. This vulnerability is especially pronounced for researchers from marginalized populations working with communities they are also part of \cite{liang2021embracing}. For these researchers, the rejection of a paper on such topics can feel deeply personal. When others dismiss the significance of these problems, it often feels like a dismissal of their own lived experiences.
\longsay{I relate to a lot of the marginalization experiences of the communities I study. When people question whether this research really matters,  it feels like they’re denying the realities I’ve lived. That kind of rejection stays with you.}{}
Researchers often felt overwhelmed or depressed when they couldn’t offer meaningful guidance to individuals facing serious safety threats. For  example, one researcher highlighted the difficulties of addressing AI-driven threats, such as the creation of manipulated or entirely fabricated explicit images. Traditional advice, such as refraining from sharing intimate photos, is increasingly ineffective given the sophistication of generative tools
Changing societal norms may be the only meaningful response, but this can feel overwhelming and unattainable in the short term.
\subsubsection{Methodological challenges}
Researchers reflected on the methodological challenges of identifying and measuring threats in HCTM.  A central challenge is that researchers often work with incomplete or imperfect data while trying to model harms that are nuanced, context-specific, and resistant to quantification. For example, much of this work relies on self-reported data, such as interviews or focus groups, which can be limited by recall issues or social desirability bias.
Researchers also noted that individuals may not fully recognize the threats they face.  For example, one researcher described how participants routinely used smart TVs without realizing they were internet-connected or capable of collecting and sharing personal data.   This made it difficult to surface accurate threat models without prompting, which  risked introducing   bias.  Even researchers struggled to understand risks when companies did not disclose how systems collect, store, or share data. Estimating threat prevalence was similarly difficult.
\longsay{Measuring prevalence is really hard --  even at [big tech company], even with all the data we have. Like, there's no one with enough of a view across the whole internet to understand what real prevalence metrics look like for lots of these things.}{}

Researchers also described facing ethical dilemmas when they became aware of threats that participants themselves had not recognized, especially when no clear solutions or mitigations were available. 
\longsay{Should a researcher inform people about the threats they face, or should they let them remain unaware? There’s a real moral dilemma in that.}{ }

Many emphasized the importance of building trust with communities, which often required slow, informal engagement: attending events, running workshops, or simply showing up without any promise of academic reward. 
Partnerships with community organizations could help support trust-building, but also introduced subtle pressures around research outcomes.
\longsay{And so when you, like, recruit through an organization, there can be this feeling of like, well, we helped you, therefore we were hoping that your results are what we wanted to hear.}{}

Establishing trust also raised tensions between allyship and objectivity. 
Researchers spoke about the   challenge of maintaining scientific rigor, particularly during data collection and threat elicitation, where personal empathy had to be carefully managed to avoid inducing bias. 
\longsay{Being as scientific and as factual as possible...  I think with people it can get very emotional, right? And you really are going to feel for for the folks that you're working with.}{Elissa}
Finally, researchers described persistent recruitment challenges. Online efforts often attracted scam accounts or bots, a problem also documented in HCI research \cite{panicker2024understanding}, but verifying participants via camera felt intrusive, especially in sensitive contexts. 
Researchers also found it difficult to retain participants in longer-term studies and, as a result, sometimes avoided multi-phase or longitudinal studies altogether. Recruitment for small or high-risk groups was slower and required more effort to convey the study’s value. In some cases, researchers turned to proxies, such as case managers instead of migrants or teachers and parents instead of children, to speak to community experiences without direct access. This may also be appropriate for reducing harm when working with at-risk populations \cite{bellini2024sok}.

\subsubsection{Impact}
\label{sec:impact_challenges}

Researchers shared that translating their findings into impact was often slow, uncertain, and structurally constrained. For those conducting qualitative or community-based work, findings were often grounded in a single population or organization, making it difficult to generalize across settings with different cultures, processes, or priorities.
\longsay{ I run into this a lot too, especially with qualitative work—it’s often a bounded case study. You’re looking at one organization, and every organization is so different. They have their own culture, their own processes, their own way of doing things. So it can be really hard to figure out how to make those findings actionable elsewhere. }{}

Several participants pointed to a disconnect between researchers and practitioners, echoing prior work on the researcher-practitioner gap in human-centered cybersecurity~\cite{haney2024towards}. Researchers often lacked access to implementation partners, and practitioners were unaware of relevant research. One participant noted they only learned about academic work after leaving industry, pointing to a broader gap between the two communities.

Participants also described technical barriers to building and maintaining consumer-facing tools. While LLM-based assistants like ChatGPT and Claude have made prototyping easier, deploying these tools at scale remained difficult. Researchers faced challenges around compatibility across systems, managing software updates, and supporting diverse user configurations.
These issues echo and add to prior experiences from long-running academic deployments, where researchers struggled with limited user engagement, and the difficulty of communicating complex behaviors without alarming users \cite{apthorpe_crowdsourcing_2020}. 

Participants in our study framed these challenges 
as structural problems tied to academic incentives, labor and funding models.
Broader impact through press, public engagement, or policy work took time and was rarely rewarded.
Similarly, for example, researchers shared that graduate students, who often led human-centered threat modeling studies, usually left after publication, making it difficult to follow through on {future work.} 
\longsay{I think that's a big tension we have in academia, which is, like, it would be great if someone did this thing, but, you know, this student wants to graduate and go, so probably not them.}{}

Participants grappled with the question of whether and when technology should be positioned as the solution to sociotechnical problems. While technical interventions were seen as potentially valuable, researchers expressed uncertainty about how to move from contextual insights to actionable designs. 
\longsay{I’m not exactly sure how to bridge that gap—there are probably useful ways that technical computer security research could contribute, but translating findings from a specific population into something actionable has turned out to be a pretty rare opportunity.}{}

Some also described feeling discouraged when designing interventions, recognizing that a solution may not be able to fully address a given harm.
They highlighted the tension between wanting comprehensive fixes and the need to build incrementally.
 \longsay{And I think sometimes we can have a little bit of the-perfect-is-the-enemy-of-the-good thing, which is like, \say{Oh, we can't build anything because it's not going to address all the things.}}{}

\subsubsection{Publishing}

Researchers described the difficulties of publishing HCTM work within traditional privacy and security venues. This work often requires slow, deep, on-the-ground community-based research (\S\ref{subsec:groundwork}) that may be best achieved with qualitative methods. 
Participants reflected on the tension between doing what was methodologically appropriate for their research and what would be accepted by reviewers. In many cases, they felt compelled to prioritize field norms, such as using large survey samples, inter-rater reliability measures, or standardized demographic categories, over practices better suited to their population or research goals. One researcher described how \textit{\say{everyone’s got their pet qualitative coding methodology,}} and they only consider that to be the \say{right} way.
For those working with small, hard-to-reach communities, these expectations frequently conflicted with what was possible or ethical.
 
Similar tensions have been observed in HCI. Soden et al. argue that interpretive research is often reviewed through positivist lenses, leading to inappropriate expectations such as frequency counts, standardized coding procedures, or participant IDs for every quote. They caution that such expectations can result in a kind of \say{methods theater,} where surface-level signals of rigor are prioritized over depth, reflexivity, and epistemological coherence \cite{soden2024evaluating}.

Beyond methods, researchers also described the pressure to demonstrate novelty and impact. Studying issues like systemic harassment or surveillance, they felt reviewers expected clear solutions, even when the problems were structural and resistant to simple fixes.

\longsay{And I'm like, if I knew how to solve gender stereotypes in society, trust me, I would have, you know.}{P5}

Researchers described the challenge of presenting human-centered findings in ways that communicate to a broad audience the substance of a human-centered threat model.   Rich, contextual data did not easily translate into diagrams that provide both substance and a meaningful overview. One researcher recounted collaborating with an applied cryptographer to create a diagram that mapped user-reported threats across devices and data flows. Despite their efforts, they eventually abandoned the visualization, unable to produce something that felt both accurate and meaningful. 

Finally,  researchers described a recurring contradiction in how their work was evaluated. Findings were at times dismissed as lacking significance and, paradoxically, also criticized as overly obvious. 
\longsay{There's this weird contradiction where some people say this research isn't important or this problem doesn't exist, but then they'll turn around and say your findings are obvious. It's really frustrating.}{P5}

\subsection{Next Steps}
\label{sec:whats-next}
HCTM has produced a wealth of insights and recommendations,  
however synthesizing these into actionable next steps remains a significant challenge.
We asked researchers about their visions for the future of human-centered threat modeling in part to understand how the field might move from surfacing threats to supporting real-world change. Several described grappling with this problem, with one calling it \textit{\say{the million dollar question.}}
Many of the strategies they described apply across security, privacy, and HCI \cite{colusso2017translational, haney2024towards}. Among our participants, there was a strong drive for impact with their HCTM work, centered in their values and motivations to help improve security and privacy outcomes for the people they worked with.

Participants stressed the need to get research in front of the people who can act on it, rather than waiting for someone to stumble across a paper in a proceedings archive. 
They encouraged researchers to go beyond academic publishing by sharing their work through public channels like op-eds, press coverage, and social media. These formats were seen as useful for drawing attention from policymakers, journalists, and platform teams. They noted that achieving change beyond academia may require hiring consultants, lab coordinators, or op-ed writers who can help sustain partnerships, communicate findings, and build long term capacity for impact.

\subsubsection{Engaging with industry}
\label{subsec:industry}

While current academic work often identifies clear implications for design and platform governance, researchers felt that these findings rarely reached the teams positioned to act on them. This reflects what Haney et al. describe as a persistent \say{research-practice gap} in human-centered cybersecurity: even when researchers produce actionable insights, they often struggle to make them visible, relevant, or timely enough for practitioner uptake \cite{haney2024towards}. Some participant suggestions  that mirror Haney et al. include (1) more intentional collaborations with platforms and companies, (2) building more structured relationships with industry by proactively identifying entry points, such as existing contacts, industry events, or invited talks; and (3) rethinking how academic work is packaged for industry, suggesting companion outputs, such as slide decks, one-pagers, or wireframes, that make findings more accessible to tech companies.

Other findings add depth and nuance to Haney et al. 
While their work focuses on engagement with security practitioners \cite{haney2024towards}, our participants emphasized the need to reach a broader range of industry stakeholders, including designers and product teams, who may not self-identify as security practitioners but whose decisions profoundly shape users’ security experiences. 

Participants also raised questions about how to measure influence when industry engagement succeeds. 
They emphasized the importance of developing more transparent and reciprocal feedback channels, so that researchers can better understand whether, when, and how their work shapes platform policy or design.
    
\longsay{We made a recommendation for {gender-based matching} and published it  in a  paper. Then, [the next year, a major platform] implemented a similar feature. I can’t say whether it was because of our work or not, but coincidentally, after we made the recommendation, the feature was implemented.}{}

Some participants saw value in pursuing a different route: developing and deploying solutions themselves when platforms   were uninterested in  certain problems. However, they also described the significant challenges involved in this path, such as maintaining compatibility across systems, testing at scale, and ensuring public visibility (\S \ref{sec:impact_challenges}). Without large support teams or marketing budgets,   researchers found it difficult to sustain these tools over time. As a result, several participants called for stronger infrastructure to support academic deployments, including funding for technical staff, long-term maintenance, and outreach.

\subsubsection{Engaging with policymaking}
\label{subsec:policy}
Because HCTM often surfaces harms that fall outside existing legal categories, participants saw value in shaping how policy problems are defined. Some began by including policy-relevant language in their papers, such as concrete recommendations for regulators.
Others engaged more directly by responding to federal requests for comment or submitting input to the Federal Register, indicating that this could get lead to invitations for deeper participation in the policymaking process.

Researchers also saw value in taking a leave from their university to work in the policy sector, and in using their findings to push back against problematic narratives or suggest alternate approaches rooted in empirical evidence. \longsay{So in the technical domain, when we see policymakers running down questionable paths, that’s usually when I end up doing some advocacy.}{}
Several participants focused on theory-building as a way to influence how policy problems are conceptualized. Because regulation often responds to defined violations, they saw value in translating human-centered risks into frameworks that policymakers could act on. 
\longsay{I think I lean more heavily on theory building [...] 
like, here is a theoretical framework for talking about risk models and privacy and security, to actionable outcomes. And because regulations are based on violations, we can transform risk into a form of violation in some way -- and then come up with things to do there. }{}

Despite strong interest in policy engagement,   participants reported having little training or support in how to do this work effectively. They called for more resources to help researchers write for policy audiences, understand regulatory processes, and build relationships with decision-makers.

\subsubsection{Engaging with society}
\label{subsec:society}

Participants envisioned a future in which HCTM expands to include deeper engagement with societal infrastructures, particularly in settings where technological interventions alone are unlikely to address root causes of harm.  

Researchers called for the field to work more intentionally with local, community-based educators, including librarians, community center staff, and other trusted figures who already support digital safety practices. While these individuals may not have formal technical backgrounds, participants emphasized that they hold valuable local knowledge and community trust. They envisioned collaborations to co-create tools, trainings, and educational materials. 
\longsay{So for example,  I’m shifting away from focusing on recommending design or policy changes. Instead, I’m partnering with these community educators who are already on the ground, working directly with people to teach them about these topics, and I’m trying to find ways to further empower them, trying to help them help other people.}{}

Participants suggested that reaching out to  journalists could be another valuable pathway for societal engagement, helping to amplify research-informed solutions and bring attention to under-addressed harms through public discourse.

Participants emphasized that engaging with society requires creating thoughtful mechanisms of societal engagement. Two exemplary projects illustrate how research can inform real-world practice: NYU's \textit{IoT Inspector}\footnote{\url{https://inspector.engineering.nyu.edu/}}  project \cite{danny2020iot} promotes public empowerment through an open-source tool that enables users to audit their smart home devices, and Cornell Tech's  \textit{Clinic to End Tech Abuse}\footnote{\url{https://ceta.tech.cornell.edu/}} embeds research within support services, providing front-line services to survivors of intimate partner violence facing technology abuse \cite{freed2017digital, freed2018stalker }.

\longsay{As a community, we need to move forward toward putting these ideas into reality.  For example, in the context of intimate partner violence, there have been clinics and similar efforts, and researchers are also working with those organizations to translate what they’ve found and explore how they can further support these populations.}{}

\subsubsection{Changes in the research community}

 Participants called for systemic changes to a research culture that values novelty and polished outcomes but rarely supports incremental progress, long-term thinking, or the messy realities of system design and maintenance.
 While HCTM has made progress in understanding user experiences and proposing design recommendations, participants expressed a desire for more actionable follow-through.
\longsay{One concern I have for the whole community  is that we put a lot of work into understanding people and on deliberate, detailed discussions about what our recommendations should be. But I feel like there’s less attention to the actual implementations, to actually building those systems, and getting them tested in the real world.}{}
They described how design work is rarely a one-time effort; it requires iteration, real-world testing, and often years of sustained work. This long, uncertain process does not align well with current academic incentives, which tend to reward polished outcomes and fast results. They called for the community to value contributions that document failure, surface obstacles, and reflect the complexities of design. 
 \longsay{System building isn’t something that can be done in one short effort and then it’s finished. It has to be a very iterative process, and the amount of work involved can take years to complete. And that kind of work often isn’t appreciated in the research community, because there’s a desire to see a successful end product}{}
 Others wanted contributions like coalition-building, educational tools, and theory-building to be recognized alongside technical solutions. Some chose to publish in venues like CSCW, where frameworks, patterns, and reflective analysis are more welcomed.

Participants called for more deliberate attention to the quality and context of design recommendations. Rather than stopping at generic suggestions like calls for more transparency, they encouraged deeper reflection on platform incentives, implementation barriers, and the broader sociotechnical dynamics shaping intervention success. Some researchers called for reexamining the implicit standards around writing design recommendations, which can lead to repeating familiar ideas even when they may not be the most effective or contextually appropriate. 
 
One researcher said they would be more willing to develop concrete recommendations if there were formal pathways for practitioners to engage with them, such as  industry feedback loops, or dedicated venues that facilitate this kind of exchange. \longsay{What I usually miss in these recommendations is the bigger picture—the context and the other factors that make it hard to implement them. Like the incentives to make money, or how bigger platforms may not want to do something if it costs money. That discussion is often missing.}{}

Participants described the need to identify broader patterns, rather than treating each population in isolation. They noted that many concerns exist across populations, so \say{\textit{suddenly it's not a small issue anymore}}.
They encouraged work that maps how certain harms and protective behaviors recur across sociotechnical contexts to surface which threats are widely shared, which vary in salience, and which tradeoffs matter most across settings. 
{Several pointed to existing works that draw such connections across fragmented research areas, including systematizations by Thomas et al. \cite{thomas2021sok}, Warford et al. \cite{warford2022sok}, and other works examining parallels across contexts such as online dating and gig work \cite{rivera2024safer}, or undocumented immigrants and incarcerated individuals \cite{owens2025understanding}.}
They called for more infrastructure grants to support data gathering that would be needed to identify emerging security and privacy issues across diverse contexts.

\label{sec:findings}

\section{Discussion}

We discuss key lessons learned across researchers’ current practices, motivations, and challenges. We then reflect on future directions for advancing HCTM based on the needs and opportunities participants surfaced.

\subsection{Lessons Learned}

\paragraph{HCTM should be a collaborative process between researchers and participants.} Our interviews demonstrate there are some limitations to conducting HCTM as an exercise that is purely devoted to understanding participant experiences. While centering humans is essential, researcher expertise can add new dimensions, particularly in areas where people may be unaware of risks they may face. Likewise, solutions to mitigating threats likely hinge on collaborations that include participant-centered design and expert perspectives.

\paragraph{Values are embedded in the work.}
The HCTM community brings together a marvelous breadth of researchers, each contributing distinct perspectives and values. Some center deep community engagement, others build infrastructure. Some pursue justice, others advocate for autonomy, privacy, or ensuring everyone has a voice.
We see value in learning to recognize the values embedded in published work, both to foster deeper collaboration and to strengthen papers through more diverse, critical perspectives.
However, explicit disclosure of values in positionality statements brings some risks \cite{liang2021embracing}; we now see a growing concern that they may also invite scrutiny or jeopardize funding,  amid the recent backlash against equity-focused research in the U.S. \cite{nyt2025lgbt, nyt2025fund}.

\paragraph{Groundwork is a continuous responsibility.}
Our findings reinforce that groundwork is not a preliminary step that can be completed and set aside. Rather, groundwork is a continuous, evolving responsibility that remains essential throughout the research process. Building trust, connecting with people to understand their perceptions, adapting methods to emerging insights, adjusting to new technological capabilities, and maintaining embedded relationships with communities are central to producing credible and ethically grounded threat models. Newer researchers, in particular, should be aware that effective HCTM cannot be achieved by \say{diving in} without sustained investment in groundwork.

\paragraph{Structural barriers limit impact, not researcher intentions.}
Our study   highlights that many barriers to impact arise not from researcher intentions, but from broader structural constraints. Short project timelines, graduation cycles, a lack of institutional mechanisms for sustaining partnerships, and limited recognition for public-facing work all limit the ability of HCTM research to produce long-term change. Those most successful at maintaining long-term engagement have built institutional support, engaged in partnerships outside the university, or developed research programs around a general-purpose infrastructure.

\paragraph{Impact must be imagined as collective and incremental, not individual and immediate.}
Building relationships, shaping norms, influencing policy, and sustaining impact requires long-term vision and coalition-building over years, rather than the completion of a single paper or technical intervention. Researchers should design studies with a five- to ten-year vision for how their work might contribute to broader social change. 

\subsection{Future Directions}

\subsubsection{Strengthening the impact of HCTM}
Participants emphasized that expanding the scope of HCTM is essential for the field’s future. Studies that result in small-scale, rich contextual insights are valuable for informing the threat landscape of particular communities or technologies. Participants likewise expressed that identifying threats or harms alone is not enough. To fulfill its goal of reducing harm, HCTM must confront the challenges of turning insights into action, whether in technology, policy, or society.  Moving from understanding to intervention is rarely straightforward. Building bridges to industry (\S\ref{subsec:industry}), engaging in policy processes (\S\ref{subsec:policy}), and collaborating with local educators and community infrastructures (\S\ref{subsec:society}) all represent promising pathways for extending impact, but each demands forms of labor, expertise, and time that are difficult to sustain under current academic structures.
There also is no standard for reporting the downstream effects of a study, the uptake of its recommendations, or the sustainability of its interventions. This limits the field’s ability to assess whether HCTM improves security outcomes or which approaches work best in different settings.

\textcolor{blue}{}{In addition to gaps in sustaining impact, our interviews suggest a need for greater communication and systematization among the researchers that practice HCTM. 
The field is mixing researchers from security, privacy, and HCI backgrounds, so there will naturally be differences in terminology, approaches, methods, and even what makes a strong contribution. 
Increased communication among researchers could improve understanding, strengthen and broaden research quality, and lead to increased collaboration. Even simply recognizing that they are addressing similar forms of harm through different lenses would be helpful.
Establishing a more consistent vocabulary and shared conceptual grounding across research areas could make it easier to compare methods, accumulate findings, and communicate their value to external stakeholders. }

Addressing these gaps will also require rethinking how human-centered security research is supported. Strengthening impact depends on interdisciplinary collaboration, partnerships beyond academia, and long-term institutional commitment. The field would benefit from more intentional, cross-sector spaces, such as policy-focused workshops or informal networks, that bring together researchers, industry partners, and policymakers to facilitate the translation of research into action. 
Additionally, we could imagine efforts for HCTM modeled after initiatives like Hackers on the Hill \cite{hackersonthehill}, where researchers and practitioners engage directly with policymakers to discuss paths for change. 
The security and privacy community would benefit from discussions around central questions, such as how human-centered research should be evaluated, how impact can be measured, and how to scale both funding and impact.

\subsubsection{Embracing pluralism of methods}
The reflections in this study suggest that HCTM must maintain openness to pluralism. Participants envisioned diverse pathways for impact, recognizing that the threats people face, the strategies for addressing them, and the sociotechnical environments in which they occur will continue to differ across contexts. Similarly, Wei et al. argue that epistemic diversity of methods is essential to uncovering the underlying reasons behind sociodemographic differences in security behaviors \cite{wei2024sok}.

This also has implications for how the community evaluates research. Reviewers must remain open to the idea that different populations may require different methods, that smaller sample sizes may sometimes be appropriate, and that different approaches to analysis may be necessary depending on the context. These reflections align with similar calls in HCI \cite{soden2024evaluating}. Some researchers have begun efforts to address challenges in reviewing diverse methods; for example, Jessica Vitak has compiled a community-shared document offering strategies for responding to common reviewer  {misunderstandings} of qualitative work \cite{vitak_qualreview}.
The community could seek partnerships in other disciplines to help strengthen the use of methods that are relatively new to computer science. The community could likewise seek collaboration between qualitative researchers with systems researchers or applied cryptographers to work jointly on problems, rather than staying in silos. Discussion among researchers with different approaches to threat modeling could push them out of comfort zones and unify effort.
The field’s growth may depend not on standardizing how threat modeling is done, but on strengthening its ability to navigate complexity, and similarly recognizing   when technical interventions are appropriate, when social or policy interventions are needed, and when translation across domains is necessary.

\subsubsection{Navigating political constraints}
Our analysis of these interviews coincides with a new political climate in the U.S., with numerous cancellations of research grants \cite{nyt2025misinfo, nyt2025lgbt, nyt2025fund},
in some cases specifically targeting work that seeks to help marginalized and at-risk populations \cite{singh2025epa}. We now face an  existential question of whether research of this nature can still be funded in the U.S. and how researchers might work within this climate to continue to advance the work.

Separate, subsequent conversations with colleagues have indicated a need to carefully word grant applications to avoid running afoul of unapproved words or ideas, a circumstance that echoes some previous periods in U.S. history \cite{steele2020history}, such as McCarthyism,
but is relatively new to this generation of researchers. This may be particularly difficult when values such as social justice are embedded in the research \cite{BhaleraoEthics} and the motivation for the research draws on a large body of community-based participatory research that focuses on marginalized or at-risk populations.
It may be easier to receive funding for HCTM work that is more technology-focused, however positioning this work within the broader HCTM research space likely still requires referencing work and ideas that are not approved by U.S. government agencies.
Thus researchers may need to use international or non-profit collaborations to find funding that enables the work to continue. This could simultaneously help broaden the field to include non-WEIRD populations \cite{hasegawa2024weird}.

\section{Conclusion}

Our findings demonstrate that threat modeling often takes shape through iterative and reflexive work rather than through a single defined process. Participants reflected on the values that motivate their work, including care, justice, and accountability, though these values sometimes come into tension with practical constraints. Participants describe a wide range of approaches to threat elicitation, many of which are developed or adapted based on the needs of their communities and the limitations of existing tools. Participants enumerated a variety of structural challenges that hinder HCTM work, such as securing long-term funding, institutional support, or clear pathways to action, which limits their ability to sustain relationships or implement findings. 
As interest in human-centered approaches to security grows, these insights inform how the research community defines good practice, allocates resources, and builds structures that allow this work to be sustained over time.

\section{Ethics considerations}
This study was approved by our institutional review board. All participants provided informed consent and were told they could skip questions or withdraw at any time. We chose not to offer compensation for participation, to avoid distorting relationships between researchers and participants who are active in the same field. To protect participant identities, we redacted information that directly identified participants from transcripts and replaced specific research topics or community names with anonymized alternatives where necessary. Before submission, we sent the paper draft and a short summary to all participants (\S\ref{sec:mem-checking}) so they could review how we represented their views and suggest clarifications. Because our study discussed researchers’ own work, which is often public, we also asked participants to flag any passages that might inadvertently identify them or their collaborators. We incorporated all requested edits to ensure participants were represented anonymously, accurately, and respectfully. We do not plan to release transcripts upon publication because the collection of stories a participant tells, along with references to their own work to explain process or illustrate a point, would likely compromise their identity.






%

\bibliographystyle{abbrv}
\bibliography{bib, corpus}
\appendices
\section{Community-centered groundwork strategies}
\label{sec:strategies}
\begin{table*}
\caption{Community-centered Groundwork Strategies}
\label{tab:groundwork}
\scriptsize
 
\centering
\renewcommand{\arraystretch}{1.3}

\begin{tabularx}{\textwidth}{p{0.2in} p{1.6in} p{2.8in} p{1.75in}}
\toprule
& \textbf{Strategy} & \textbf{Example   Researcher Story} & \textbf{CBPR Principles  } \\
\midrule

\multirow{4}{*}{\raisebox{-1.5\height}{\rotatebox[origin=c]{90}{\scriptsize \textbf{Connecting}}}} 
    & Gain entry through community insiders 
    & A researcher gained access to a closed, sensitive online group after being introduced by a trusted member who vouched for them. 
    & Community as unit of identity;  build on community strengths; equitable partnerships \\

    & \cellcolor{gray!15}Partner with local organizations and unions 
    & \cellcolor{gray!15}One researcher spent the first few months of their postdoc meeting with community organizations to understand local concerns before defining their research problem. 
    & \cellcolor{gray!15} Build on community strengths \\

    & Run pre-research workshops 
    & A participant conducted security and privacy workshops with teens for over a year before starting formal study procedures. 
    &  Co-learning and capacity building; mutual benefits;   sustainable, long-term partnerships; equitable partnerships \\

    & \cellcolor{gray!15}Leverage informal personal networks 
    & \cellcolor{gray!15}A participant connected with a community through a friend of a friend within the usable security field who was part of that community. 
    & \cellcolor{gray!15}Community as unit of identity;   Build on community strengths \\

\midrule

\multirow{5}{*}{\raisebox{-1.7\height}{\rotatebox[origin=c]{90}{\scriptsize \textbf{Understanding}}}} 
    & Read community forums and online discourse 
    & Before designing their studies, researchers spent hours on Reddit reading personal narratives shared by community members to learn how communities discussed harm and support-seeking. 
    & Focus  on local issues of public concern; build on
community strengths \\

    & \cellcolor{gray!15}Follow popular and interdisciplinary discourse 
    & \cellcolor{gray!15}One participant learned from Instagram infographics about race, gender, and class to inform their understanding of how safety threats were framed and experienced by the community. 
    & \cellcolor{gray!15}Focus  on local issues of public concern  \\

    & Partner with experts 
    & A research team studying children’s digital safety partnered with pediatricians and child psychologists to help design age-appropriate and ethically sound methods. 
    & Co-learning and capacity building;   equitable partnerships \\

    & \cellcolor{gray!15}Draw on researcher positionality 
    & \cellcolor{gray!15}A participant reflected on how their own experiences of marginalization informed the way they asked questions and engaged with participants. 
    & \cellcolor{gray!15}Co-learning and capacity building \\

    & Learn the community’s language and framing 
    & Researchers intentionally listened for the vocabulary people used to describe harm and adapted their own terminology and framing accordingly. 
    & Community as unit of identity;   focus  on local issues of public concern   \\

\midrule

\multirow{3}{*}{\raisebox{-1.5\height}{\rotatebox[origin=c]{90}{\scriptsize \textbf{Working}}}} 
    & Hire community insiders as consultants 
    & One researcher hired a paid consultant from the community they were studying to help review and refine recruitment materials and study instruments. 
    &  Build on community strengths;   equitable partnerships \\

    & \cellcolor{gray!15}Collaborate with other researchers from the community 
    & \cellcolor{gray!15}A participant collaborated with a co-author from the same region as their study participants to help interpret cultural and linguistic nuances in the data. 
    & \cellcolor{gray!15} Equitable partnerships; co-learning and capacity building \\

    & Check your findings with the community (Member checking) 
    & A researcher working with a disabled community periodically shared preliminary findings throughout the study to gather feedback and ensure the results accurately reflected participants’ lived experiences. 
    & Share results with all partners \\

\bottomrule
\end{tabularx}
\end{table*}

Table~\ref{tab:groundwork} presents the strategies researchers described and the CBPR principles they reflect.
\section{Interview Guide}
\begin{itemize}
  \item Intro, consent, purpose of the study
  \begin{itemize}

    \item First, do you think, broadly, my characterization of your work is correct?
   
  \end{itemize}
 \item How do you see yourself positioning your research within the broader landscape of human-centered threat 
    modeling?
 
  \item In your work on human-centered threat modeling, how do you build understanding of the people or users whose experiences inform your study?
  \begin{itemize}
    \item What steps do you take to ground your work in their perspectives, especially when you are not part of that context yourself?
  \end{itemize}

  \item What kinds of challenges did you encounter while conducting this research?
  \begin{itemize}
    \item How did you navigate or respond to those challenges?
    \item Despite the difficulties, what do you find most meaningful or rewarding about doing this kind of work?
  \end{itemize}

 \item How did you surface or elicit the threats relevant to the people or users you studied?
  \begin{itemize}
    \item Did you ask directly about threats and harms, or did you find it more effective to use a different, perhaps more nuanced, approach to uncover these insights?
 
  \end{itemize}

  \item What are your long-term goals with this line of research?
  \begin{itemize}
    \item In what ways do you hope your work will support or benefit the communities or users you study?
    \item What strategies do you use to move toward those goals?
  \end{itemize}

  \item How do you see your research making an impact?
  \begin{itemize}
    \item How do you assess whether your work is reaching the people or systems you hope to influence?
  \end{itemize}

  \item We now have a growing body of research with recommendations that touch on technical systems, social practices, and policy change.
  \begin{itemize}
    \item From your perspective, what should be next for our community?
    \item How do we move from recommendations to meaningful action?
  \end{itemize}

  \item What personal values guide your work in human-centered threat modeling?
  \begin{itemize}
    \item Do these values shape the kinds of questions you ask, the methods you use, or how you engage with the community or users your research involves?
    \item Has your thinking about threat modeling evolved over time?
    \begin{itemize}
      \item Were there moments in your research where your own assumptions or methods shifted based on what you learned?
    \end{itemize}
  \end{itemize}

  \item Based on your experience, what advice would you give to other researchers working in this field?
  \begin{itemize}
    \item Are there particular practices or approaches you’ve found especially helpful?
   
  \end{itemize}

  \item Thank you for sharing your insights and experiences.
  
    \item Before we close, do you have any final thoughts, questions, or anything you feel was important but we didn’t cover?

\end{itemize}
\section{Member Checking}
\label{member_checking}
Dear [Participant Name],

We hope this message finds you well. Thank you again for generously sharing your time and insights for our study on human-centered threat modeling.  We’ve put together a draft of the paper based on the interviews, and we’re reaching out to check whether the way we’ve represented your perspective (and the field more broadly) feels accurate to you before we move forward with submission.

We’ve attached both (1) a summary of findings if you prefer a high-level overview and (2) the full paper draft in case you’d like to see the complete write-up and how any anonymous quotes were used in support of the findings. You are welcome to read either or both.

If you’re willing, we’d appreciate your thoughts on any part of the paper, but particularly on the following questions:
\begin{enumerate}

    \item After reading through the findings, what are your general thoughts?

    \item How accurately do you feel the findings captured your thoughts/experiences?

    \item Is there anything you feel should be added or emphasized more to better represent your experience?

    \item Is there anything you would prefer be removed or revised? If so, could you share why?

\end{enumerate}

Any feedback you’re able to provide is entirely optional but would be incredibly valuable. Please feel free to respond in whatever format is most convenient (e.g., send us the PDF with your annotated comments, reply to this email). Thank you again for contributing to this work!
\end{document}